\documentclass{iau}
\usepackage{graphicx,xspace,hyperref,natbib}

\newcommand{\Gaia}{\textit{Gaia}\xspace}
\newcommand{\kms}{km\,s$^{-1}$\xspace}
\newcommand{\masyr}{mas\,yr$^{-1}$\xspace}

\title[Using \Gaia for studying Milky Way star clusters]
{Using \Gaia for studying\\ Milky Way star clusters}

\author[Eugene Vasiliev]{Eugene Vasiliev$^{1,2}$}

\affiliation{$^1$Institute of Astronomy, University of Cambridge, UK\\
$^2$Lebedev Physical Institute, Moscow, Russia\\
email: {\tt eugvas@lpi.ru}
}

\pubyear{2019}
\volume{351}  
\jname{Star Clusters: From the Milky Way to the Early Universe}
\editors{A. Bragaglia, M.B. Davies, A. Sills \& E. Vesperini, eds.}
\begin{document}

\maketitle

\begin{abstract}
We review the implications of the \Gaia Data Release 2 catalogue for studying the dynamics of  Milky Way globular clusters, focusing on two separate topics.\\
The first one is the analysis of the full 6-dimensional phase-space distribution of the entire population of Milky Way globular clusters: their mean proper motions (PM) can be measured with an exquisite precision (down to 0.05~\masyr, including systematic errors). Using these data, and a suitable ansatz for the steady-state distribution function (DF) of the cluster population, we then determine simultaneously the best-fit parameters of this DF and the total Milky Way potential.
We also discuss possible correlated structures in the space of integrals of motion.\\
The second topic addresses the internal dynamics of a few dozen of the closest and richest globular clusters, again using the Gaia PM to measure the velocity dispersion and internal rotation, with a proper treatment of spatially correlated systematic errors. Clear rotation signatures are detected in 10 clusters, and a few more show weaker signatures at a level $\gtrsim 0.05$~\masyr. PM dispersion profiles can be reliably measured down to 0.1~\masyr, and agree well with the line-of-sight velocity dispersion profiles from the literature.
\keywords{proper motions -- globular clusters: general -- Galaxy: kinematics and dynamics}
\end{abstract}

\firstsection
\section{Introduction}

The second data release (DR2) of the \Gaia mission \citep{Brown2018} provides proper motions (PM) and broad-band photometry for more than a billion stars down to magnitude $G=21$, based on 22 months of observations. Despite some limitations (e.g., no special treatment of binary stars, and poor completeness in crowded fields), this dataset is a giant leap forward for studies of Milky Way and its constituents, and has been used in more than 1000 papers over the last year.

The mean PM of roughly half of the Milky Way globular clusters have been calculated in \citet{Helmi2018a}. Subsequently, mean PM of almost the entire population of the $\sim 150$ clusters \citep{Harris1996} have been independently measured by \citet{Vasiliev2019b} and \citet{Baumgardt2019}. The former study additionally considered the constraints on the Milky Way potential from the dynamics of globular clusters, while the latter also analyzed the internal PM for most of them. \citet{Vasiliev2019c} further extended the analysis of internal kinematics by considering the impact of spatially correlated systematic errors in \Gaia astrometry.

\section{Membership determination}

The first step in analysis is the selection of cluster members. We use only the positions and PM of stars in a circular area centered on each cluster, but not their magnitudes or colors, which may be affected by extinction and (in the case of colors) crowding. We select only stars with parallax values consistent with the inverse distance to a cluster at $3\sigma$ level, satisfying certain quality criteria.

Instead of using hard cuts (e.g. removing $3\sigma$ outliers in PM), we employ a probabilistic Gaussian mixture classification method, 
described in the appendix of \citet{Vasiliev2019b}; the code is provided at \url{https://github.com/GalacticDynamics-Oxford/GaiaTools}.
First, we define a two-component model for the joint distribution of stars in the sky plane and PM space.
The density of cluster stars is assumed to follow a particular functional form (Plummer profile) in the sky plane, and their PM have an isotropic Gaussian distribution with a spatially-varying width $\sigma_\mathrm{c}(R)$, while the distribution of field stars is uniform on the sky plane and is described by a generic 2d Gaussian in the PM space. The fraction of cluster stars, their mean PM and the dispersion $\sigma_\mathrm{c}(0)$, the scale radius of the Plummer profile, the mean and covariance of the Gaussian distribution of field stars, are all free parameters in the model. 
The likelihood of measuring a particular value of PM is given by the sum of the values of cluster and field distribution functions (DFs), convolved with measurement errors (different for each star).
The parameters of the model are then optimized to maximize the log-likelihood of the entire dataset, which is a sum of log-likelihoods of individual stars.
Finally, after the best-fit parameters have been determined, we evaluate the membership probability for each star.

\section{Systematic errors in astrometry}

\begin{figure}[t]
\includegraphics[width=2.68in]{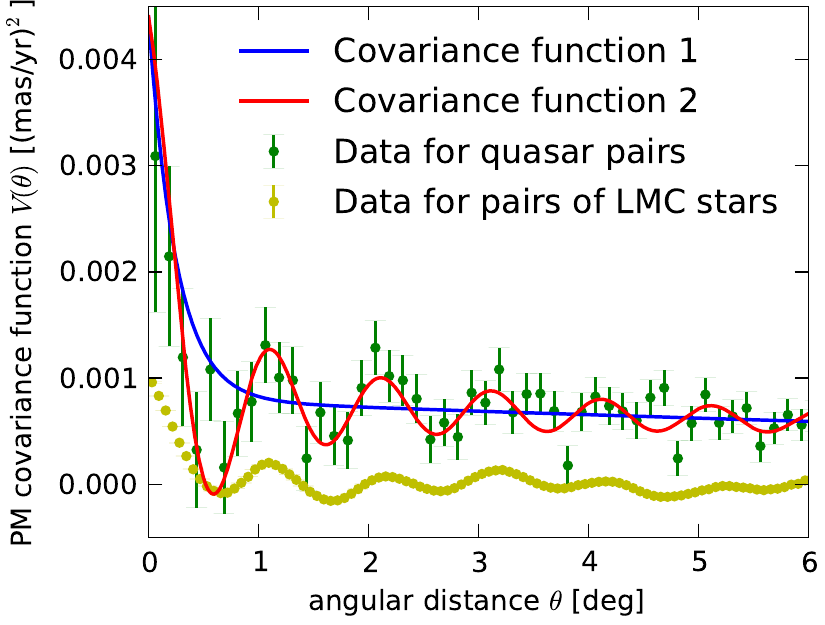} 
\includegraphics[width=2.60in]{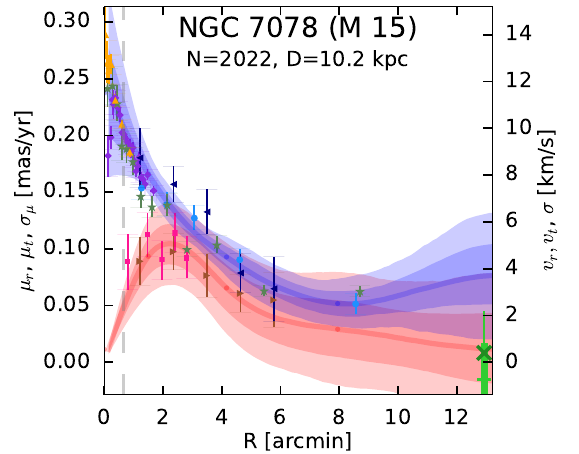}
\caption{Left panel: correlation function of PM measurement errors as a function of angular separation between two sources.\protect\\
Right panel: example of measurement of internal kinematics for the globular cluster NGC 7078. 
Shaded regions show the 68\% and 95\% confidence intervals for the rotation (red) and PM dispersion (blue) profiles, determined in \citet{Vasiliev2019c}; points with error bars show measurements from other studies.
} \label{fig:covfnc_and_pm_profiles}
\end{figure}

Because of the way that \Gaia surveys the sky, measurement errors appear to be spatially correlated on angular scales $\lesssim 0.5^\circ$. In the first approximation, the two-point correlation function depends only on the angular distance between sources (Figure~15 in \citealt{Lindegren2018}, or left panel of Figure~\ref{fig:covfnc_and_pm_profiles}), and a suitable functional form is given by Equation~1 in \citet{Vasiliev2019c}. The latter paper introduces the mathematical formalism for taking into account these spatially correlated PM errors when performing model fits (in particular, measuring the mean PM of a cluster, or its internal kinematics -- rotation and velocity dispersion). 
The uncertainties on the mean PM of almost all clusters are dominated by the systematic errors, and are at the level $0.05-0.1$~\masyr.

\section{Internal kinematics of globular clusters}

We use this formalism to fit more sophisticated models, in which the PM of cluster stars has a radially-varying rotational component and a radially-varying [isotropic] dispersion. By calibrating the method on mock datasets with full account of spatially correlated systematic errors, we find that the rotation can be reliably measured when its peak amplitude exceeds $\sim 0.05$~\masyr, and the PM dispersion when it exceeds $\sim 0.1$~\masyr; these conditions are fulfilled for $\sim 60$ clusters mostly within 10~kpc and having at least 100 member stars passing all quality cuts.

We find high-significance rotation signatures in 8 clusters, and a weaker signatures in another 10 clusters; the results agree well with independent studies of \citet{Bianchini2018,Sollima2019,Jindal2019}, even though none of these papers took into account systematic errors. The PM dispersion profiles for 60 clusters are also in a good agreement with independent PM measurements of \citet{Baumgardt2019} based on \Gaia data, \textit{HST}-based PM measurements \citep{Watkins2015} in the central parts of 22 clusters, and various studies of line-of-sight velocity dispersion profiles, although for several most crowded clusters our PM  dispersions are likely unreliable in the central parts. An example is shown in the right panel of Figure~\ref{fig:covfnc_and_pm_profiles}.

\section{Dynamics of the entire Milky Way globular cluster system}

\begin{figure}[t]
\includegraphics[width=5.3in]{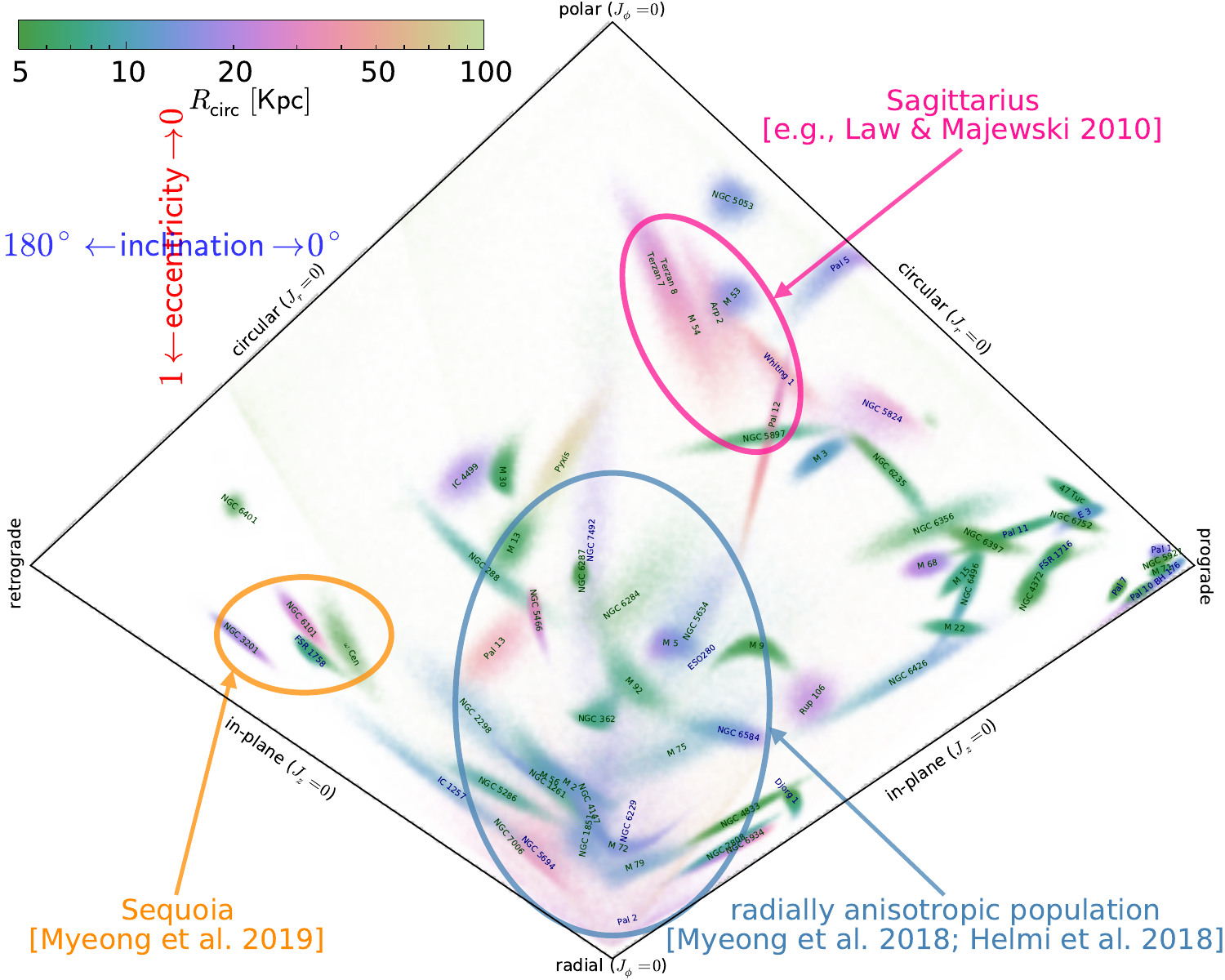} 
\caption{
Action-space map of the outer Milky Way globular clusters (a modified version of Figure~5 of \citealt{Vasiliev2019b}). The horizontal and vertical coordinates show the actions $J_\phi$ and $J_z-J_r$, normalized by the angular momentum $L_\mathrm{circ}(E)$ of a circular orbit with the given energy; they roughly correspond to eccentricity and inclination. The color shows the energy, expressed in terms of the radius of a circular orbit. Individual clusters are shown by clouds of Monte Carlo samples representing the measurement uncertainties. Encircled are possible distinct kinematic subgroups identified by \citet{Law2010,Myeong2018,Myeong2019,Helmi2018b}.
} \label{fig:actions}
\end{figure}

With our measurements of mean PM of clusters, and the distances and line-of-sight velocities taken from the literature, we analyze the distribution of [almost] the entire population of Milky Way globular clusters in the 6d phase space. We confirm the previously known overall rotation of clusters (especially metal-rich ones) in the inner part ($r\lesssim 10$~kpc) of the Milky Way, with the mean $v_\phi$ reaching $\sim 70 - 80$~\kms. The velocity dispersion is nearly isotropic ($\sim 100$~\kms) in the inner part, and becomes more radially anisotropic further out.
The distribution of clusters in the space of integrals of motion (actions) reveals several possible kinematic subgroups (Figure~\ref{fig:actions}), which may be remnants of ancient accretion events.

We use clusters as dynamical probes of the Milky Way potential. Assuming a particular functional form for the DF of clusters in the action space (similar to \citealt{Binney2017} and \citealt{Posti2019}), we explore the range of parameters of the Milky Way potential and the DF consistent with the measured 6d phase-space coordinates of clusters (again convolved with observational uncertainties, which are mostly dominated by distance errors).
We find an approximately flat rotation curve ($v_\mathrm{circ} \simeq 200-240$~\kms between 10 and 100~kpc), consistent with the best-fit potential of \citet{McMillan2017}, but higher than the one from \citet{Bovy2015}. The enclosed mass within 50 kpc is inferred to be $(0.45-0.65)\times10^{12}\;M_\odot$, and within 100 kpc -- $(0.65-1.2)\times10^{12}\;M_\odot$; this is mostly consistent with other recent measurements (e.g., \citealt{Watkins2019}), but higher than the results of \citet{Eadie2019} based on the same kinematic sample.

\vspace*{5mm}
This work uses the data from the European Space Agency mission \Gaia (\url{https://www.cosmos.esa.int/gaia}).
This work was supported by the European Research council under the 7th Framework programme (grant No.\ 308024). 


\end{document}